\documentclass[a4paper,11pt]{article}
\pdfoutput=1

\usepackage{cite}
\DeclareRobustCommand{\e}{\epsilon}

\begin{document}
\unitlength1cm
\begin{titlepage}
\vspace*{-1cm}
\begin{flushright}
ZU-TH 09/21
\end{flushright}
\vskip 3.5cm

\begin{center}
{\Large\bf The three-loop singlet contribution\\[2mm] to the massless axial-vector quark form factor}
\vskip 1.cm
{\large  T.~Gehrmann} and {\large A.~Primo} \vskip .7cm
{\it Physik-Institut, Universit\"at Z\"urich,
Winterthurerstrasse 190,\\ CH-8057 Z\"urich, Switzerland}

\end{center}
\vskip 2cm

\begin{abstract}
We compute the three-loop corrections to the quark axial vector form factor in massless QCD, focusing on the 
pure-singlet contributions where the axial vector current couples to a closed quark loop. Employing the Larin prescription for 
$\gamma_5$, we discuss the UV renormalization of the form factor. The infrared singularity structure of the resulting 
singlet axial-vector form factor is explained from infrared factorization, defining a finite remainder function.
\end{abstract}
\end{titlepage}
\newpage

The quark form factors describe the coupling of a quark-antiquark pair to an external current, which can be a vector, scalar, axial-vector  
or pseudo-scalar. Higher order perturbative corrections to these form factors provide important universal 
information~\cite{Mueller:1979ih,Sen:1981sd,Magnea:1990zb} on 
anomalous dimensions, 
and they constitute the purely virtual corrections to important collider processes such as gauge boson production, 
Higgs boson decay or deeply inelastic scattering. After renormalization of ultraviolet (UV) divergences, the form factors remain divergent due 
to infrared (IR) poles. 

For massless quarks, the three-loop QCD corrections to vector~\cite{Moch:2005id,Baikov:2009bg,Gehrmann:2010ue}, scalar~\cite{Gehrmann:2014vha}
and pseudo-scalar~\cite{Ahmed:2015qpa} form factors were derived in the literature, and 
important progress has been made recently towards the four-loop corrections to the vector form 
factor~\cite{Henn:2019swt,Lee:2019zop,vonManteuffel:2020vjv,Agarwal:2021zft}. 
All massive form factors are known to two-loop order in 
QCD~\cite{Bernreuther:2004ih,Bernreuther:2004th,Bernreuther:2005rw,Bernreuther:2005gw}, supplemented by partial three-loop 
results~\cite{Blumlein:2019oas}. 

Owing to chirality conservation, the massless vector and axial-vector form factors can differ from each other only through contributions 
where the external current couples to a closed quark loop, which is then connected to the external quark-antiquark pair through 
virtual gluon exchanges. These so-called pure-singlet contributions (PS) occur for the first time at two loops, where they vanish for the vector form factor
while yielding an IR-finite contribution for the axial-vector form factor~\cite{Bernreuther:2005rw}. 
The three-loop pure-singlet contributions to the vector form factor were computed earlier~\cite{Baikov:2009bg,Gehrmann:2010ue} 
and found to be IR-finite. All contributions where the axial vector current insertion couples to the external quarks are denoted as non-singlet (NS), and the 
sum of non-singlet and pure singlet contributions yields the singlet (S) form factor.  
In the present letter, 
we derive the three-loop pure singlet contributions to the axial-vector form factor in massless QCD, thereby completing the full set of massless 
three-loop quark form factors. 

Using dimensional regularization to handle both infrared and ultraviolet divergences, one must extend the 
four-dimensional chirality projection operator $\gamma_5$ to symbolic $d=4-2\epsilon$ space-time dimensions. For this purpose, we follow the 
prescription introduced by Larin~\cite{Larin:1993tq}, 
which replaces the symmetrized axial vector vertex factor as follows:
\begin{equation}
\gamma_\mu\gamma_5  \to \frac{1}{2} \left( \gamma_\mu\gamma_5 - \gamma_5\gamma_\mu\right) \to \frac{i}{6} \epsilon_{\mu\nu_1\nu_2\nu_3}
\gamma^{\nu_1}\gamma^{\nu_2}\gamma^{\nu_3}\,.
\label{eq:larin}
\end{equation}
It is based on the original $\gamma_5$ formulation of 
 t'Hooft and Veltman~\cite{tHooft:1972tcz}, which was further refined by Breitenlohner and Maison~\cite{Breitenlohner:1977hr}, 
 but has the further advantage that the Lorentz 
 index space does not need to be split into $4$-dimensional and $(d-4)$-dimensional subspaces, thereby allowing all Lorentz algebra to be performed 
 in $d$ dimensions throughout. In the Larin scheme, a finite renormalization of the axial vector current is required 
 (besides the conventional UV-renormalization) to restore chirality conservation of massless quarks and to ensure the validity of 
 the chiral anomaly. The axial renormalization constants in the Larin scheme were known to three-loop order 
 for the non-singlet contributions for a long time~\cite{Larin:1993tq,Larin:1991tj},  while the finite contribution to the renormalization of the singlet 
 current has been derived only most recently~\cite{Ahmed:2021spj}. 

The axial vector vertex function is obtained by the insertion of  an off-shell axial-vector current with virtuality 
$q^2 = (p_1+p_2)^2$ between a quark-antiquark pair with on-shell momenta $p_1$ and $p_2$, yielding the Born-level expression
\begin{equation}
\bar{u}(p_2) \Gamma_{A,0}^{\mu} u(p_1) = \bar{u}(p_2)  \gamma^\mu\gamma_5  u(p_1)\,.
\end{equation}
The axial vector form factor 
is then obtained by applying a projection operator on the all-order vertex function $ \Gamma_{A}^{\mu}$:
\begin{equation}
{\cal F}_A = -\frac{3}{4(1+\epsilon)(1-\epsilon)(1-2\epsilon)(3-2\epsilon) q^2}\, {\mathrm Tr} \left( p_2 \!\!\!\! / \, \Gamma_A^\mu p_1 \!\!\!\! / \,  \gamma_\mu\gamma_5 \right)\, , \label{eq:projq}
\end{equation}
where the Larin prescription (\ref{eq:larin}) is to be applied throughout. With the above normalization, the Born-level axial vector form factor 
is equal to unity: ${\cal F}_{A}^{(0)} = 1$. The amplitude-level calculations closely follow those 
of the three-loop vector form factor~\cite{Gehrmann:2010ue}. The bare form factor is expanded in the bare QCD coupling constant 
$a_B = \alpha_{s,B}/(4\pi)$ as 
follows: 
\begin{equation}
{\cal F}_A^B = \sum_{i=0}^\infty a_B^i  {\cal F}_A^{B,(i)}\;.
\end{equation}

The $\overline{{\rm MS}}$ 
renormalization of the axial vector vertex function $\Gamma_A^\mu$ involves~\cite{Larin:1993tq,Larin:1991tj,Ahmed:2021spj} 
the renormalization of the coupling constant $Z_g$ and of the 
axial vector current insertion $Z_5^{ms}Z_5^f$, where $Z_5^{ms}$ and $Z_5^f$ denote the divergent and finite parts of the axial vector 
renormalization constant. They depend on the prescription used for $\gamma_5$ in dimensional regularization. 
Different axial vector renormalization constants are required for the singlet and non-singlet axial vector form factors. They have been computed 
to three-loop order in the Larin scheme for the non-singlet~\cite{Larin:1993tq,Larin:1991tj} and singlet~\cite{Ahmed:2021spj} axial vector current. 
Their expansion in the renormalized QCD coupling constant $a=\alpha_s(\mu^2)/(4\pi)$ reads:
\begin{equation}
Z = \sum_{i=0}^\infty a^i  Z^{(i)}\;.
\end{equation}

After renormalization, the massless non-singlet form factors for axial vector and vector  agree with each other in 
their finite parts that are obtained by subtracting their universal infrared pole structure~\cite{Catani:1998bh,Moch:2005id,Gardi:2009qi,Becher:2009cu},  
as required by chirality conservation for 
massless fermions and as obtained for a naively anti-commuting $\gamma_5$. To extract the pure singlet axial vector form factor, one takes the 
difference of the renormalized singlet and non-singlet axial vector form factors:
\begin{equation}
{\cal F}_{A,{\rm PS}} 
= {\cal F}_{A,{\rm S}} - {\cal F}_{A,{\rm NS}} \, . 
\label{eq:defPS}
\end{equation}
By taking the difference of the singlet and non-singlet renormalization constants~\cite{Larin:1993tq,Larin:1991tj,Ahmed:2021spj}, 
we define pure-singlet renormalization constants
which turn out to be useful in arranging~\cite{Bernreuther:2005rw} the different contributions in terms of non-singlet and pure-singlet type:
\begin{eqnarray}
Z_{5,{\rm PS}}^{ms} &= &Z_{5,{\rm S}}^{ms}  - Z_{5,{\rm NS}}^{ms}  \\
& = & C_F\, N_{F,J} \left( \frac{3}{\e} a^2 +\frac{(-66+109\e)C_A - 54 \e C_F + (12+2\e) N_F}{9\e^2} a^3   \right) + {\cal O}(a^4)
  \;,\nonumber \\
Z_{5,{\rm PS}}^{f} &= &Z_{5,{\rm S}}^{f}  - Z_{5,{\rm NS}}^{f}  \\
& = &  C_F\, N_{F,J} \left(  \frac{3}{2}a^2 + \frac{(-326+1404\zeta_3) C_A   + (621-1296\zeta_3)  C_F + 176 N_F}{54} a^3   \right) 
+ {\cal O}(a^4) \;, \nonumber
\end{eqnarray}
where $C_A=N$, $C_F=(N^2-1)/(2N)$ are the QCD colour factors. 
The overall power of $N_{F,J}$ can be identified with the number of quark flavours that couple to the external axial vector current, 
while $N_F$ is the number of massless quark flavours. In the following, we take $N_{F,J}=1$ throughout.

Expanding out the pure singlet axial vector form factor (\ref{eq:defPS}) in powers of the renormalized QCD coupling constant, 
\begin{equation}
{\cal F}_{A,{\rm PS}} = a^2 {\cal F}_{A,{\rm PS}}^{(2)} +  a^3 {\cal F}_{A,{\rm PS}}^{(3)} +
{\cal O}(a^4) \,,
\end{equation}
one finds the following
expressions at two and three loops:
\begin{eqnarray}
{\cal F}_{A,{\rm PS}}^{(2)}  & = &  {\cal F}_{A,{\rm PS}}^{B,(2)} +  Z_{5,{\rm PS}}^{ms,(2)}+Z_{5,{\rm PS}}^{f,(2)}\;,  \\
{\cal F}_{A,{\rm PS}}^{(3)}  & = & {\cal F}_{A,{\rm PS}}^{B,(3)} + \left( Z_{5,{\rm NS}}^{ms,(1)} -\frac{2\beta_0}{\e} \right)  
{\cal F}_{A,{\rm PS}}^{B,(2)} + 
\left( Z_{5,{\rm PS}}^{ms,(2)}+Z_{5,{\rm PS}}^{f,(2)}\right)  {\cal F}_{A,{\rm NS}}^{B,(1)}  \nonumber \\
&& +  Z_{5,{\rm NS}}^{f,(1)}  Z_{5,{\rm PS}}^{ms,(2)}+
Z_{5,{\rm PS}}^{ms,(3)}+Z_{5,{\rm PS}}^{f,(3)} \;,
\end{eqnarray}
where $\beta_0 = (11C_A/3-2N_F/3)$. All bare form factors  are computed using the Larin prescription throughout. 

The two-loop contribution has been computed previously. We reproduce the result of~\cite{Bernreuther:2005rw} and provide 
higher order terms in the $\epsilon$ expansion, as required for the study of the infrared singularity structure at higher loop orders:
\begin{eqnarray}
{\cal F}_{A,{\rm PS}}^{(2)}  &=&  C_F\Bigg[ -18 + 6 L_\mu + \frac{2\pi^2}{3} + \e \left(
          - \frac{345}{4}
          + 3 L_\mu
          + 6 L_\mu^2
          + 4 \zeta_3
          + \frac{59\pi^2}{18}
\right) \nonumber \\ &&
+ \e^2 \left( 
          - \frac{2579}{8}
          + 3 L_\mu^2
          + 4 L_\mu^3
          + \frac{146}{3}\zeta_3
          + \frac{1469\pi^2}{108}
          + \frac{4\pi^4}{45}
\right) + {\cal O} (\e^3)
\Bigg] \;,
\end{eqnarray}
where we have introduced $L_\mu = \log(-q^2/\mu^2)$. It should be noted that only the finite term in the above expression is independent on 
the prescription used for $\gamma_5$, while all  $\epsilon$-type terms are specific to the Larin scheme.

The three-loop pure singlet axial vector form factor is our main new result. It reads
\begin{eqnarray}
{\cal F}_{A,{\rm PS}}^{(3)}  &=&  \frac{1}{\e^2} C_F^2 \left(  36
          - 12 L_\mu
          - \frac{4\pi^2}{3} \right) + \frac{1}{\e} C_F^2 \left( 
           \frac{453}{2}
          - 24 L_\mu
          - 12 L_\mu^2
          - 8 \zeta_3
          - \frac{77\pi^2}{9}
          \right)
          \nonumber \\ &&
         + C_F^2\left( 
          1108
          - 75L_\mu
          + \pi^2 L_\mu
          - 24 L_\mu^2
          - 8 L_\mu^3
          - \frac{328}{3}\zeta_3
          - \frac{1225\pi^2}{27}
          - \frac{3\pi^4}{5}
\right)
          \nonumber \\ &&
 + C_F N_F \left(
          \frac{469}{9}
          - \frac{76}{3} L_\mu
          + \frac{8\pi^2}{9} L_\mu
          + 4 L_\mu^2
          - \frac{40\pi^2}{27}
         \right)
                   \nonumber \\ &&
+ C_FC_A \left(
          - \frac{7403}{18}
          + \frac{538}{3} L_\mu
          - \frac{44\pi^2}{9}L_\mu
          - 22 L_\mu^2
          + 62 \zeta_3
          + \frac{385\pi^2}{27}
          - \frac{17\pi^4}{90}
          \right)
\;.\label{eq:F3ren}
\end{eqnarray}
The divergent contributions in the above expression are of infrared origin. They can be expressed by applying the one-loop infrared singularity 
operator~\cite{Catani:1998bh}
\begin{equation}
\mathrm{I}^{(1)}_{q\bar q} = - C_F \left(\frac{2}{\e^2}+\frac{3}{\e} \right) \left(1-\e^2\frac{\pi^2}{12}\right)\;,
\end{equation} such that 
\begin{eqnarray}
{\cal F}_{A,{\rm PS}}^{(3,{\rm finite})} &=& {\cal F}_{A,{\rm PS}}^{(3)}  - \mathrm{I}^{(1)}_{q\bar q} {\cal F}_{A,{\rm PS}}^{(2)}  \\
& = & C_F^2\left( 
           \frac{409}{2}
          - 66 L_\mu
          - \frac{16\pi^2}{3}
          - \frac{8\pi^4}{15} \right) 
 \nonumber \\ &&
 + C_F N_F \left(
           \frac{469}{9}
          - \frac{76}{3} L_\mu
          + \frac{8\pi^2}{9} L_\mu
          + 4 L_\mu^2
          - \frac{40\pi^2}{27}
          \right)
           \nonumber \\ &&
+ C_F C_A \left(
         - \frac{7403}{18}
          + \frac{538}{3}L_\mu
          - \frac{44\pi^2}{9} L_\mu
          - 22 L_\mu^2
          + 62 \zeta_3
          + \frac{385\pi^2}{27}
          - \frac{17\pi^4}{90}
          \right)
          \label{eq:F3finite}
\end{eqnarray}
is finite. The finite remainder functions that are 
obtained from infrared pole subtraction~\cite{Catani:1998bh,Gardi:2009qi,Becher:2009cu}
were previously observed~\cite{Weinzierl:2011uz} to be independent on the prescription that is being used to define  
internal and external polarization states in dimensional regularization. It can therefore be expected that (\ref{eq:F3finite}) is 
independent on the $\gamma_5$-scheme, while (\ref{eq:F3ren}) is valid only in the Larin scheme.  

We observe that
the finite pure-singlet contribution of the three-loop vector form factor~\cite{Baikov:2009bg,Gehrmann:2010ue} contains terms proportional to 
$\zeta_5$, which are not present in  (\ref{eq:F3ren}) or (\ref{eq:F3finite}). The finite parts of the three-loop non-singlet form factors even contain $\pi^6$, 
which are absent in their pure-singlet counterparts. This remarkable lowering of transcendental weight deserves further study. 

The three-loop pure-singlet 
axial vector form factor contributes to coefficient functions for observables in polarized hadron collisions and 
to the three-loop coefficient function for hadronic $Z$-boson production. In the latter, its contribution cancels for 
mass-degenerate quark isospin doublets in the loop,
 such that a sizable effect can be expected only from the third-generation quarks. For these, the massless 
form factor computed here provides a reliable description of the bottom  quark contribution, while the top quark contribution remains to be derived. 

In this letter, we completed the calculation of the three-loop quark form factors in massless QCD by deriving the pure-singlet axial 
vector form factor at this order. Our computation employed the Larin scheme for $\gamma_5$ throughout in all amplitudes and projectors. 
The three-loop  pure-singlet axial 
vector form factor is infrared-divergent. Its singularity structure in accordance with the expectation from infrared factorization, such that 
a  finite remainder function can be defined. 

\section*{Acknowledgements}
This research was supported    by the Swiss National Science Foundation (SNF) under contract 200020-175595.

\end{document}